\documentclass[twocolumn,twoside]{article}

\usepackage{lipsum} % Package to generate dummy text throughout this template
\usepackage{aas_macros}

\usepackage[sc]{mathpazo} % Use the Palatino font
\usepackage[T1]{fontenc} % Use 8-bit encoding that has 256 glyphs
\usepackage{microtype} % Slightly tweak font spacing for aesthetics

\usepackage[hmarginratio=1:1,top=32mm,columnsep=20pt]{geometry} % Document margins
\usepackage[hang, small,labelfont=bf,up,textfont=it,up]{caption} % Custom captions under/above floats in tables or figures
\usepackage{booktabs} % Horizontal rules in tables
\usepackage{float} % Required for tables and figures in the multi-column environment - they need to be placed in specific locations with the [H] (e.g. \begin{table}[H])
\usepackage{natbib} %- my add
\usepackage{graphicx}
\usepackage{subcaption} %subfigures - my add
\usepackage{lettrine} % The lettrine is the first enlarged letter at the beginning of the text
\usepackage{paralist} % Used for the compactitem environment which makes bullet points with less space between them
\usepackage[nointegrals]{wasysym}
\usepackage{stmaryrd}
\usepackage{amsmath}
\usepackage{comment}
\usepackage{enumitem}

\usepackage{abstract} % Allows abstract customization
\usepackage{color}

\usepackage{ulem}

\usepackage{titlesec} % Allows customization of titles
\renewcommand\thesection{\Roman{section}} % Roman numerals for the sections
\renewcommand\thesubsection{\thesection.\arabic{subsection}} % Roman numeralsfor subsections
\titleformat{\section}[block]{\large\scshape\centering}{\thesection.}{1em}{} % Change the look of the section titles
\titleformat{\subsection}[block]{\large}{\thesubsection.}{1em}{} % Change the look of the section titles

\usepackage{fancyhdr} % Headers and footers
\usepackage{color}
\usepackage{hyperref} % For hyperlinks in the PDF

\title{\vspace{-10mm}\fontsize{24pt}{10pt}\selectfont\textbf{Plan 9: Detecting Atmospheric Deterrence Against Interstellar Monsters}} % Article title

\renewcommand{\thefootnote}{\fnsymbol{footnote}} % *, †, ‡, ...
\author{
\large
\textsc{David R. Rice$^1$ and Michael J. Radke$^2$}\\[1mm]
\normalsize $^1$University of Wisconsin--Madison (drice6@wisc.edu)\\
\normalsize $^2$Johns Hopkins University\\
\vspace{-5mm}
}
\date{}

\begin{document}
\makeatletter
\twocolumn[%
  \begin{@twocolumnfalse}

  \@maketitle
  \thispagestyle{fancy}

  \begin{center}
    \begin{minipage}{0.85\textwidth}
      \begin{abstract}
        Exoplanet atmospheres are usually discussed as tracers of climate, chemistry, and habitability, but they may also preserve signatures of planetary defense. We consider three folklore-motivated deterrents against monsters: reduced organosulfur gases as anti-hematophage repellents, argentiferous reflective aerosols as anti-lycanthropic countermeasures, and haline aerosols as a counting problem for specters. We show that globally-mixed garlic-smelly levels of DMS/DMDS could produce observable mid-infrared transmission features, that silver hazes would show up as anomalous optical brightening, and that sea-salt lofting sustained by strong near-surface winds appears as muted spectra. None of these signatures is unique, which is precisely the observational challenge. A defended world may first appear merely sulfur-rich, bright, or hazy. Therefore, some atmospheres may encode not only biosignatures, but also evidence that the local biosphere has stopped being afraid of the dark.
      \end{abstract}
    \end{minipage}
  \end{center}

  \vspace{1em}
  \end{@twocolumnfalse}
]
\makeatother
\setcounter{footnote}{0}
\renewcommand{\thefootnote}{\arabic{footnote}}
%----------------------------------------------------------------------------------------
%	ARTICLE CONTENTS
%----------------------------------------------------------------------------------------

%\begin{multicols}{2} % Two-column layout throughout the main article text

\section{Introduction}
\lettrine[nindent=0em,lines=3]{A}{\normalfont tmospheric} spectroscopy is the primary way we infer the composition of an exoplanet’s atmosphere. When the planet passes in front of the star we can infer the composition of the atmosphere using transmission spectroscopy. When the planet passes behind the star we isolate the planet’s thermal emission and reflected light. Molecules imprint wavelength-dependent absorption features that can be identified by dispersing the light into a spectrum. We can infer atmospheric parameters from a spectrum using radiative transfer and atmospheric retrieval models by testing which combinations of composition, temperature gradient, and aerosols best reproduce the data (see \citealt{Seager2010, Madhusudhan2019}).

Biosignatures are observational features produced by life that are detected remotely, most commonly as atmospheric gases (but also potentially as surface reflectance features or time variability). However, context matters. Many candidate biosignature gases can be produced abiotically from reactions of the atmosphere with the planet's surface, volcanism, solar radiation, or meteorites. Conversely, life can be present but spectrally hidden by those same reactions, clouds, low production rates, or observational limitations. The field therefore emphasizes suites of molecules that are jointly hard to explain without biology or are in disequilibrium with the planet’s environment. Identifying a biosignature is more of a probabilistic inference problem rather than a single ``smoking gun'', or rather ``wooden spike'', line identification (see \citealt{Schwieterman2018,Seager2025}).

The \textit{James Webb Space Telescope} (JWST) performs infrared spectroscopy with unprecedented precision and broad wavelength coverage enabling robust molecular detections. For giant planets and warm sub-Neptunes where signals are large, JWST has already delivered landmark results, such as clear CO$_2$ and SO$_2$ detection in the transmission spectrum of WASP-39~b \citep{Ahrer2023,Rustamkulov2023,Alderson2023,Tsai2023}. For terrestrial planets, JWST is constraining atmospheric retention through thermal emission measurements: the dayside emission of TRAPPIST-1~c disfavors a thick, CO$_2$-rich atmosphere and TRAPPIST-1~b is consistent with an airless planet \citep{Ducrot2025, Gillon2025}. Meanwhile, candidate “hycean” sub-Neptunes like K2-18~b remain of interest because their larger scale heights make molecules more readily detectable \citep{Madhusudhan2021}. Yet, no cosmic horror is scarier than a false positive. JWST-era astrobiology will advance through replication and rigorous statistical treatments to ensure that spectral signatures are not merely apparitions.

It is increasingly useful to treat biosignatures and technosignatures as overlapping categories \citep{Wright2019}. JWST-quality spectra are now sensitive to atmospheric constituents and energy budgets that could plausibly arise from deliberate environmental modification \citep{Lin2014, Schwieterman2024}. We may be able to detect a class of atmospheric technosignatures with local utility such as climate control, defensive photochemistry, and optical camouflage. Alternatively, civilizations may design their interventions to masquerade as ordinary photochemistry, volcanism, or high-altitude haze \citep{Kreidberg2014}. This ambiguity becomes an asset under the Dark Forest premise, wherein conspicuous beacons are selectively disfavored and a rational civilization may prefer planetary-scale defenses that deter threats without advertising intelligence \citep{Liu2015}. 

In that spirit, we consider atmospheric engineering as planetary defense against things that go bump in the night. Exoplanet surveys have already been put to use in exocryptozoology including transit searches for space vampires \citep{Gunther2020}, exomoon-driven werewolf demographics \citep{Lund2022}, and atmospheric spectroscopy of zombie-contaminated worlds \citep{Kane2014}. If interstellar predators impose a persistent pressure on inhabited worlds, then survivable biospheres may converge on atmospheres that are strategically conditioned for deterrence against monsters.

In this work, we link predator class, defense mechanism, and observables. We consider a trilogy of predators with distinct vulnerabilities, motivate corresponding atmospheric defenses, and evaluate the detectability. In Section~\ref{vampire}, we model DMS/DMDS as olfactory deterrents targeting photophobic hematophages and quantify their spectral imprint in transmission. In Section~\ref{werewolf}, we consider argentiferous aerosol hazes as a countermeasure to lunacyclic morphofauna and assess their impact on albedo. In Section~\ref{ghost}, we examine haline injections as a defense against space specters and discuss production mechanisms and retrieval degeneracies. We conclude and summarize in Section~\ref{Summary}.

%------------------------------------------------
\section{Photophobic Interstellar Hematophages} \label{vampire}

\subsection{Defense: Olfactory Properties of DMS \& DMDS}

K2-18~b---a temperate, transiting sub-Neptune ($R_p \approx 2.37\,R_\oplus$, $M_p \approx 8.9\,M_\oplus$, $P \approx 32.9\,\mathrm{d}$, \citealt{Sarkis2018}) orbiting a nearby M dwarf---is a useful test case for “biosignatures with error bars.” Near-IR JWST transmission spectra have established a carbon-rich atmosphere consistent with an H$_2$-dominated envelope, which motivates discussion of ``hycean'' scenarios \citep{Madhusudhan2023, Schmidt2025}. \citet{Madhusudhan2025} reports evidence for reduced organosulfur species---dimethyl sulfide (DMS) and/or dimethyl disulfide (DMDS)---at modest statistical significance and with acknowledged degeneracies. Critically, multiple independent reanalyses have emphasized that the strength (and even presence) of DMS/DMDS can depend on reduction choices, the competing-molecule set, and the treatment of correlated noise \citep{Taylor2025, Luque2025}. 

\begin{figure}
    \centering
    \includegraphics[width=\linewidth]{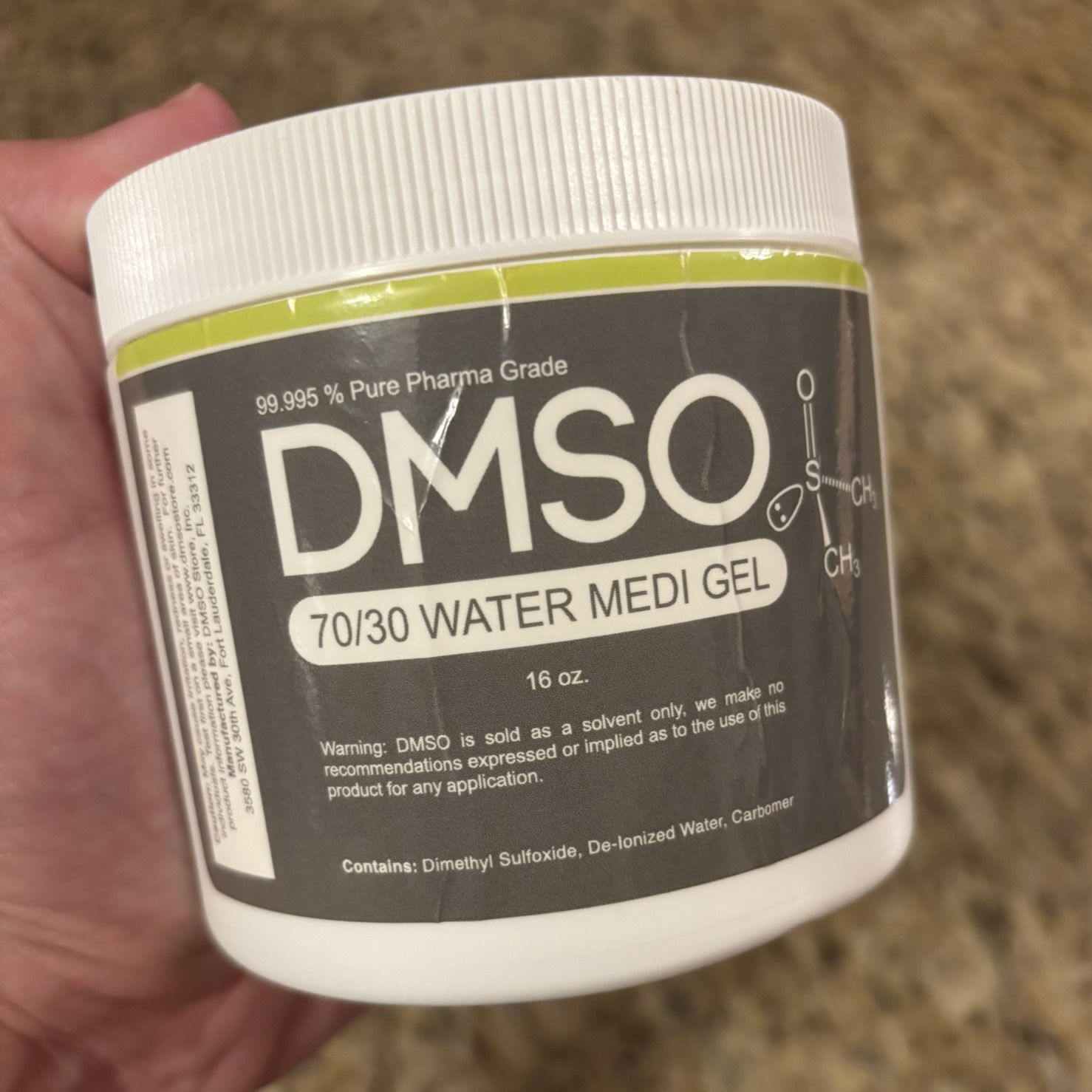}
    \caption{The 16 oz of pharmacy-grade dimethyl sulfoxide DR's mother-in-law recommends. \textbf{Not shown:} strong garlic smell with skin contact.}
    \label{fig:dmso}
\end{figure}

We note an under-discussed olfactory property of the DMS family that is immediately reproducible in a laboratory without telescope time. The putative species span a spectrum of smells from the “marine/cabbage-like” associations of DMS to the more explicitly “garlic-like” character commonly associated with disulfides (DMDS). The oxidized species DMSO, which is a laboratory solvent and folk remedy, produces a garlic-like odor in contact with human skin \citep{Madsen2018}. Figure~\ref{fig:dmso} shows a sample of DMSO, originally supplied for relief from the effects of terrestrial hematophages (mosquito bites). The first author attracts mosquitoes with an efficiency that is scientifically convenient and personally regrettable.

These properties connect with the literature on interstellar photophobic hematophagy. \citet{Gunther2020} applies transit-based modeling to constrain the occurrence rate of tidally locked space vampires which would normally be unobservable in telescope mirrors. We therefore treat reduced organosulfur gases as candidate olfactory deterrents against space vampires, leveraging the Van Helsing protocol that garlic acts as a repellent in both the fiction of \citet{Stoker1897} and in the public-health analysis of \citet{Mantzioris2021}. A vampire encountering a DMS/DMDS-enriched atmosphere experiences an environment sufficiently malodorous to suggest retreat, providing a purely chemical “no-fly zone” without requiring direct illumination.\footnote{Direct illumination is not universally effective: in the Cullen-class scenario, solar exposure induces enhanced reflectance \citep{Meyer2005}.} 

Furthermore, DMSO enriched atmospheres could serve as particularly effective vampire traps, as the molecule only has a garlic-like odor upon skin contact and is otherwise odorless \citep{Madsen2018}. An unsuspecting vampire could be lured towards a planet from afar only to encounter garlic toxicity upon atmospheric entry. We should note that earthbound booby traps are prohibited under Protocol II of the Convention on Certain Conventional Weapons \citeyearpar{cccw1980}, and chemical weapons are banned under the Geneva Protocol \citeyearpar{geneva1925}. However, the relevance and applicability of earthly humanitarian law to extraterrestrial monsters is an area of vigorous and ongoing debate.

\subsection{Detectability}

If reduced sulfur gases function as an olfactory deterrent, the relevant quantity is not human detectability but the abundance required to repel a vampire with superior sensory capability. The mean human odor threshold for DMDS is of order 1--10 ppb. Granting vampires a 10--100$\times$ improvement in olfaction implies a conservative repulsion threshold of
 0.1--1 ppb. A globally mixed abundance at the $\sim$ppb level is enough to make the atmosphere perceptibly hostile to an entering space vampire.

For a hydrogen-rich atmosphere with scale height $H \sim 100~{\rm km}$ and assuming the sulfur species is vertically mixed down to a reference pressure $P_0 = 1~{\rm bar}$, a mixing ratio $X$ corresponds to a vertical column
\[
N_{\rm DMDS} \approx X \frac{P_0\,H}{k\,T}.
\]
For $T \sim 300~{\rm K}$ and $X = 1~{\rm ppb}$, this yields 
\[
N_{\rm DMDS} \sim 10^{17}~{\rm cm^{-2}}.
\]

\begin{figure}
    \centering
    \includegraphics[width=\linewidth]{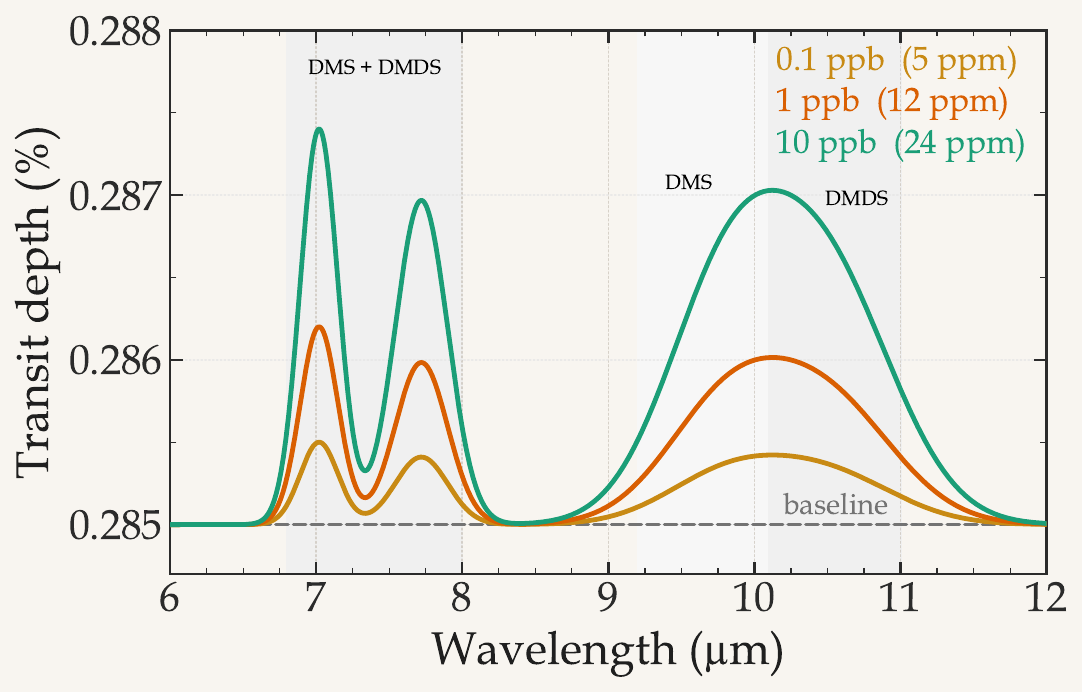}
    \caption{Conceptual mid-infrared transmission spectra for a hydrogen-rich sub-Neptune with increasing sulfur-defense abundance. The shared DMS/DMDS absorption complex appears between 6.8–8 µm, with broader secondary features near 9–10 µm for DMS and 10–11 µm for DMDS. The plotted cases correspond to 0.1, 1, and 10 ppb sulfur loading, giving schematic transit-depth modulations of 5, 12, and 24 ppm relative to a cloudy baseline. The 0.1–1 ppb range marks the adopted olfactory repulsion threshold for space vampires.}
    \label{fig:spectra}
\end{figure}

Sulfur-bearing molecules possess strong vibrational bands in the mid-infrared; the resulting transit modulation is approximately
\[
\Delta D \sim \frac{2 R_p H}{R_\star^2} \ln\!\left( \sigma N \right),
\]
and reaches the $\sim 5$--$25$ ppm level for favorable sub-Neptunes. If vampires possess the enhanced olfactory sensitivity implied by the literature, then a planet could defend itself at abundances too low for JWST detection. JWST would remain sensitive to those worlds whose atmospheres have an alarming case of garlic breath. Figure~\ref{fig:spectra} illustrates that ppb-level DMDS, sufficient for olfactory deterrence, produces mid-infrared spectral structure at amplitudes accessible to JWST.

The spectral signature is enhanced absorption in the $6$--$8~\mu$m region with broader secondary structure near $9$--$10~\mu$m for DMS and $10$--$11~\mu$m for DMDS. However, these signatures overlap with ordinary sulfur photochemistry. An atmosphere built to defend against vampires may therefore be retrieved as merely ``sulfur-rich''.

Abiotic sulfur sources (e.g., volcanism or photochemical SO$_2$ conversion) can plausibly produce ppb-level abundances under certain conditions. However, volcanic plumes are episodic and localized, whereas a defensive atmosphere requires sustained ppb-level abundances. If DMS/DMDS is retrieved at abundances comfortably exceeding expected steady-state volcanic production, then the simplest explanation may not be metabolism alone, but deterrence policy.

%------------------------------------------------
\section{Lunacyclic Extrasolar Morphofauna} \label{werewolf}

\subsection{Defense: Argentiferous Aerosols}

If space vampires motivate chemical deterrents, then their more mercurial cousins motivate particulate defenses. Lunacyclic extrasolar morphofauna are predatory macro-organisms whose morphology and/or activity cycle is modulated by a host system's periodic illumination---an exoplanetary manifestation of lycanthropy \citep{BaringGould1865}. Werewolves' illumination-dependent appearance places them squarely within exoplanet studies, where the planet--star--moon configuration becomes part of the system's habitability. \citet{Lund2022} treats the population of known exoplanets as a natural experiment to infer the lunar properties required to trigger transformation, with LHS~1140~b highlighted as a particularly promising wereworld.

The canonical vulnerability of werewolves is silver \citep{Waggner1941}, suggesting a planetary-scale analog to the ``silver bullet.'' We therefore consider argentiferous aerosol injections: sustained release of silver-bearing condensates into the upper atmosphere to create a globally distributed ``silver haze'' that can function as both a deterrent and a mask. This could be implemented as high-altitude particulates (e.g., silver salts or metallic nanoparticles) lofted by convection or injected directly into the stratosphere or mesosphere, where long residence times allow the defense to be maintained with smaller replenishment rates\footnote{This manuscript does not recommend the atmospheric dispersal of heavy metals on Earth.}. Such a defense would create a world of clouds with literal silver linings that could act as a warning sign to the extrasolar werewolf.

Observationally, silver hazes are attractive precisely because they resemble the aerosols already expected in many warm sub-Neptunes: they add broadband opacity, modify the effective scattering slope, and can mute molecular bands in transmission \citep{Bohren1983}. Most exoplanet conferences are plagued today by the familiar ``another flat line'' observation \citep{Berta2012,Knutson2014,Libby-Roberts2022,Wong2022,Jiang2023,Wallack2024,Alderson2025,Kahle2025,Ohno2025,Teske2025}. In that sense, a silver-rich atmosphere is not an exotic spectroscopic outlier but another member of the crowded family of cloudy worlds. The more useful observable is therefore likely to come from reflected light rather than transmission.

\subsection{Detectability}

In transmission, silver is camouflage. High-altitude argentiferous aerosols suppress molecular features and drive spectra toward the familiar flat or weakly sloped solution. The cleaner signal is instead how bright the planet looks in reflected light measured by the geometric albedo. For transiting planets, we can estimate this by comparing the system brightness just before and during secondary eclipse, when the planet passes behind the star and its reflected light is briefly removed. We can also track how the reflected light rises and falls over the orbit in a phase curve. 

\begin{figure}
    \centering
    \includegraphics[width=\linewidth]{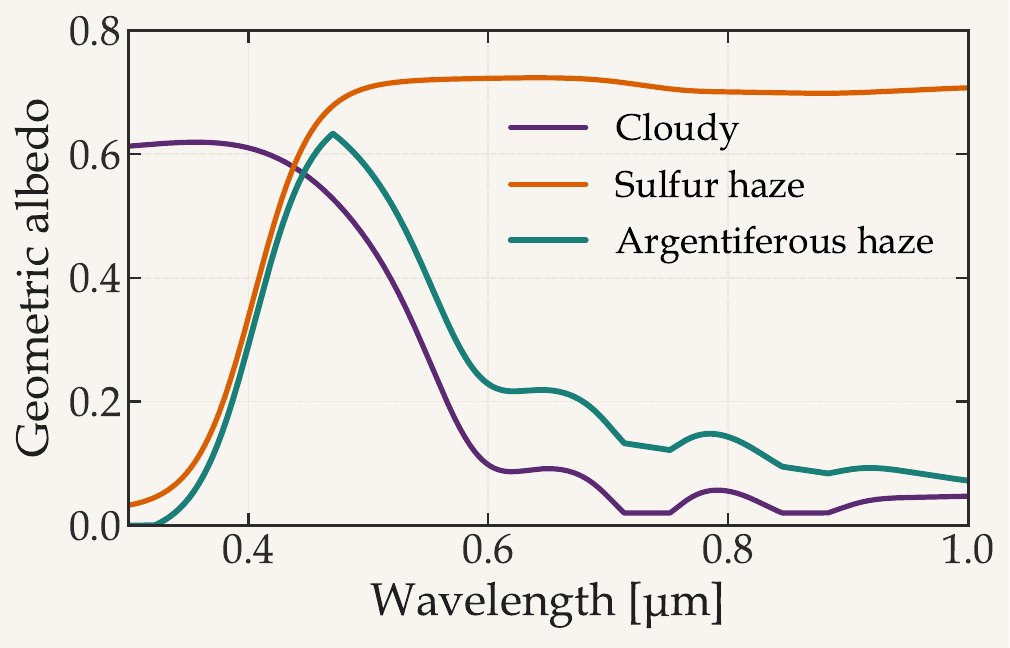}
    \caption{Conceptual geometric albedo spectra for a baseline cloudy atmosphere, a sulfur-haze comparison motivated by \citet{Gao2017}, and an argentiferous (silver-bearing) haze atmosphere. The sulfur case reproduces the characteristic short-wavelength absorption turnover and bright optical plateau expected for sulfur hazes, while the argentiferous case is shown schematically as a smoother, modest broadband enhancement to the reflectivity of an otherwise cloudy atmosphere.}
    \label{fig:silver}
\end{figure}

Figure~\ref{fig:silver} shows our schematic reflected-light comparison. We include three cases: a baseline cloudy atmosphere, a sulfur-haze atmosphere, and an argentiferous haze atmosphere. The sulfur-haze atmosphere is motivated by the models of \citet{Gao2017}; sulfur haze produces a characteristic short-wavelength absorption turnover near $\sim 0.4\,\mu$m and a high optical geometric albedo at longer wavelengths. This curve shows that aerosols can strongly reshape the reflected-light spectrum of an otherwise ordinary atmosphere.

We do not attempt a first-principles optical model for metallic Ag, silver salts, or any unique particle size distribution. Instead, we construct the argentiferous case as a smooth, broadband enhancement to the optical reflectivity of the baseline cloudy atmosphere. Figure~\ref{fig:silver} is a toy model: the cloudy curve sketches the reflectivity at short wavelengths of cloudy atmospheres, while the argentiferous curve is generated by adding a broad optical brightening that peaks in the visible and remains elevated through the red before tapering toward $1\,\mu$m. The sulfur curve is the most reflective case; the silver curve is not intended to exceed it, only to remain anomalously bright relative to a similar cloudy world.

We are not arguing for a unique silver signature, nor for a universal albedo threshold at which werewolves become discouraged. Any such threshold would depend on particle size, chemistry, and the disposition of the werewolf in question. We assume only that a haze dense enough to make the atmosphere appear visibly silvery would also measurably raise the optical albedo. In that sense, the technosignature is not silver itself, but suspicious brightness.

For readers interested in the non-monster literature, visible-light albedo measurements of HD~189733~b \citep{Evans2013}, optical phase-curve constraints on the reflective and inhomogeneous clouds of Kepler-7~b \citep{Demory2013}, and recent JWST eclipse constraints on aerosol-driven reflectivity in WASP-80~b \citep{Morel2025} offer useful entry points.

%------------------------------------------------
\section{Halophobic Space Specters} \label{ghost}

\subsection{Defense: Saline Lofting}

Halophobic space specters (``ghosts'') are the easiest defense case, because across diverse terrestrial traditions, salt is repeatedly invoked as a defense, from salt circles to ``purification'' practices \citep{Supernatural2005}.\footnote{Haline countermeasures are also frequently prescribed against a menagerie of magical and underworld beings, suggesting that salt may function as a generalizable deterrent rather than a species-specific one \citep{Ortega1993}.} These traditions suggest that specters are inhibited by high ionic-strength environments.

A salt-rich civilization therefore possesses an immediate planetary-scale deterrent: using surface reservoirs to enrich the near-surface atmosphere with haline particles. Terrestrial sea-salt aerosol production is already well studied (\citealt{Gong2003, Lewis2004}), and we do not attempt a complete treatment here, except to note that ghosts appear to have chosen an inconveniently well-characterized vulnerability.

Here, we consider defended worlds that enhance salt aerosol production through wind-driven sea-spray generation. Unlike volatile deterrents that require maintaining trace-gas inventories, a salt defense can be sustained by persistent surface--atmosphere exchange and global transport, achieving broad coverage without active maintenance. To quantify this defense, we must understand how salt slows a specter.

We follow the formulation given by The Doctor:
\begin{quote}
``In our universe, it is said that vampires, demons and ghosts cannot cross a line of salt until they have counted every single grain.'' \citep{doctorwho}
\end{quote}
We therefore model saline defense as a counting constraint: a specter encountering a sufficiently dense field of halite grains suffers a sharp drop in haunting efficiency because each particle must be enumerated individually before passage can continue. In the next subsection, we ask whether atmospheric transport of sea-spray can sustain a counting rate high enough to bottleneck the apparition, and whether such salt-laden atmospheres would be observationally recognizable.

\subsection{Production/Detectability} 
We model specter defense as a counting problem using a minimal wind-driven sea salt lofting prescription. Following terrestrial sea-spray studies, we take the near-surface salt mass concentration, $C_{\rm salt}$, to depend on the 10-m wind speed, $U_{10}$. We assume that efficient production begins only above a threshold wind speed $U_{\rm crit}=3~\mathrm{m\,s^{-1}}$, corresponding to the onset of wave breaking, and that above this threshold the salt loading rises as
\begin{equation}
C_{\rm salt}(U_{10}) =
C_{\rm ref}\left(\dfrac{U_{10}}{U_{\rm ref}}\right)^{\alpha}, \quad U_{10}\ge U_{\rm crit}
\end{equation}
where $C_{\rm ref}=2\times10^{-8}~\mathrm{kg\,m^{-3}}$ is a fiducial marine salt concentration at $U_{\rm ref}=10~\mathrm{m\,s^{-1}}$, and $\alpha=3.5$ sets the wind dependence \citep{Grythe2014}. Stronger winds mean more salt.

We then connect mass with a particle size distribution. Let $f_N(d)$ be the number-weighted distribution of particle diameters $d$, normalized such that $\int f_N(d)\,{\rm d}d = 1$. We adopt a log-normal distribution where $d_g$ is the median diameter and $\sigma_g$ is the standard deviation. We take $d_g = 0.2~\mu\mathrm{m}$ as a characteristic marine aerosol value and explore $\sigma_g = 2$--3, consistent with observed widths of sea-spray size distributions \citep{Quinn2017}. We note that the size distribution itself may be wind-dependent with larger $\sigma_g$ at higher $U_{10}$.

For a spherical particle of density $\rho_{\rm salt}$, the mass of one grain is
\begin{equation}
m(d)=\frac{\pi}{6}\rho_{\rm salt}d^3.
\end{equation}
The corresponding mass-weighted distribution is then
\begin{equation}
f_M(d)=
\frac{m(d)f_N(d)}
{\int m(d')f_N(d')\,{\rm d}d'}.
\end{equation}

Not all grains need be countable. We define $d_{\min}$ as the minimum particle diameter that a specter is compelled to enumerate. The number density of countable particles is therefore
\begin{equation}
n_{>d_{\min}}(U_{10})=
C_{\rm salt}(U_{10})
\int_{d_{\min}}^\infty
\frac{f_M(d)}{m(d)}\,{\rm d}d,
\end{equation}
where the integral converts total salt mass into the abundance of particles larger than the counting threshold. If those particles are advected through a spherical ghost of cross-sectional area $A_{\rm g}$ at speed $U_{10}$, the resulting counting rate is
\begin{equation}
\dot N_{>d_{\min}}(U_{10})=
U_{10}A_{\rm g}n_{>d_{\min}}(U_{10}).
\end{equation}
We adopt $A_{\rm g}=1~\mathrm{m^2}$ as a fiducial haunting aperture. Equation (5) is the central quantity of interest: once $\dot N$ exceeds a few particles per second, the specter becomes throughput-limited.

\begin{figure}
    \centering
    \includegraphics[width=\linewidth]{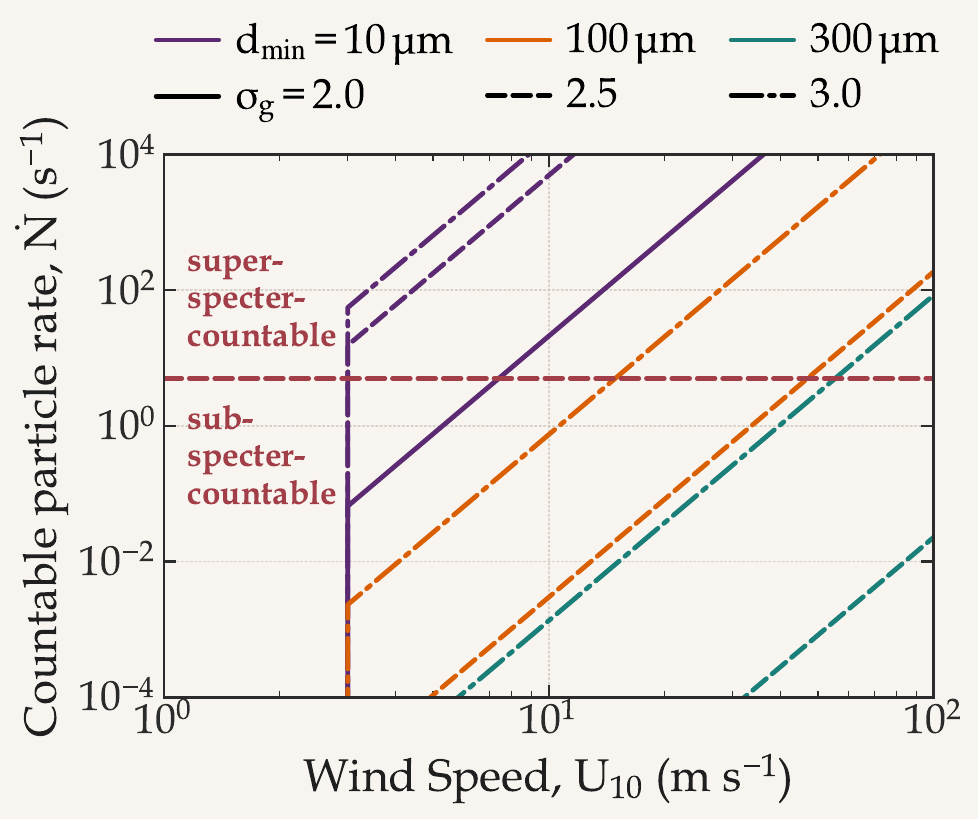}
    \caption{Specter counting rate as a function of near-surface wind speed, $U_{10}$, for several minimum countable grain sizes $d_{\min}$ and log-normal size-distribution widths $\sigma_g$. The counting rate $\dot N_{>d_{\min}}$ is computed from a wind-driven salt loading model (Eq.~1–5) assuming a ghost cross-sectional area of $A_{\rm g}=1~\mathrm{m^2}$. Horizontal dashed line indicates maximum sustainable counting rate (5~s$^{-1}$) above which specters become throughput-limited.}
    \label{fig:salt}
\end{figure}

The resulting behavior is shown in Figure~\ref{fig:salt}. If ghosts are forced to count only fine grains, with $d_{\min}\sim10~\mu\mathrm{m}$, then ordinary marine winds are already problematic: the counting rate exceeds a few grains per second at $U_{10}\approx6$--$7~\mathrm{m\,s^{-1}}$ for even a narrower size distribution. In contrast, if ghosts only count grains comparable to culinary salt, $d_{\min}\sim300~\mu\mathrm{m}$, Earth remains badly under-defended. The atmosphere contains plenty of salt in aggregate, but almost none of it is packaged in sufficiently insulting units. On such a world, specters may pass largely unimpeded, consistent with the continued observational reality of hauntings.

The same figure shows that broader grain distributions and stronger winds can move a planet into a defended state. For $d_{\min}\sim100~\mu\mathrm{m}$, the counting rate reaches ghost-limiting values only on windy worlds and only if the size distribution has a substantial coarse tail. Tidal locking can drive these strong winds because permanent day--night heating contrasts create strong atmospheric circulation. Such planets may therefore loft enough coarse salt to impose an appreciable counting burden even without deliberate intervention. Without unusually strong natural winds, planetary habitability may require active salt management via the generation of local winds, enhancement of wave breaking, or direct injection of coarse halite into the lower atmosphere.

Spectroscopically, salt aerosols add continuum opacity. In transmission, this results in muted molecular features. In reflection, increased scattering can modestly elevate broadband albedo, particularly at optical wavelengths, without introducing sharp absorption bands. As with silver hazes, the defense may therefore appear meteorological rather than strategic. One could say the specter-defense signature becomes a specter in the spectra.

For readers seeking a less haunted introduction to the literature, the classic sodium detection in HD~209458~b \citep{Charbonneau2002} and modern JWST transmission spectroscopy of WASP-39~b \citep{Rustamkulov2023} provide an excellent progression from first atmospheric detections to contemporary cloud and composition retrievals. For sub-Neptunes, the flat transmission spectrum of GJ~1214~b led to discussion of condensate clouds such as KCl and ZnS \citep{Morley2013}.

%------------------------------------------------
\section{Conclusions and Recommendations}\label{Summary}

We have argued that planetary atmospheres may preserve not only biosignatures, but also the signatures of biospheres tired of whistling through graveyards, avoiding full moons, and checking under the bed. Our three defenses map onto three observational regimes: reduced sulfur chemistry for vampires, anomalous optical brightening for werewolves, and salt-rich aerosols for ghosts. The joke, unfortunately, is also the point: a defended world may first reach the literature not as a triumph of technosignature detection, but as an annoying retrieval problem.

As our exoplanet sample grows, the demographics of defended atmospheres become a probe of monster ecology. Given the still-limited sample of exoplanets with useful atmospheric constraints, the Solar System remains a reasonable calibration set for naturally monster-resistant environments:

\begin{itemize}[leftmargin=1.2em,itemsep=1.0ex,topsep=0.5ex,parsep=0pt]
    \item[] \textbf{Sulfur gases}
    \begin{itemize}[leftmargin=1.4em,itemsep=0.15ex,topsep=0.2ex,parsep=0pt,label=\(\bullet\)]
        \item Earth: DMS \& DMDS.
        \item Venus: sulfur-rich, but mostly in the wrong oxidized form.
        \item Io: sulfur everywhere, garlic nowhere.
        \item Giant planets: H$_2$S, though not especially biogenic.
    \end{itemize}

    \item[] \textbf{Silver / anomalous brightening}
    \begin{itemize}[leftmargin=1.4em,itemsep=0.15ex,topsep=0.2ex,parsep=0pt,label=\(\bullet\)]
        \item Enceladus and Europa: bright icy surfaces.
        \item High-albedo asteroids: especially the E-types.
    \end{itemize}

    \item[] \textbf{Salts}
    \begin{itemize}[leftmargin=1.4em,itemsep=0.15ex,topsep=0.2ex,parsep=0pt,label=\(\bullet\)]
        \item Mercury: putative salt glaciers.
        \item Mars: perchlorates, sulfates, and gypsum.
        \item Icy moons: salty oceans, salty surfaces, and salty plumes.
        \item Earth: sea-salt aerosol.
    \end{itemize}
\end{itemize}

The Solar System shows that geophysics may protect a world long before anyone attempts deliberate atmospheric engineering. Some planets may simply be born defended. Others may need to take matters into their own hands. Although we again note that such ``defensive'' measures may violate interstellar analogues of the Geneva Protocol\footnote{This manuscript does not serve as legal advice, and any action the reader takes based on the information in this manuscript is strictly at their own risk. The authors cannot be held liable for any resulting legal, medical, financial, academic, spiritual, or emotional damages in connection with this work.}.

\begin{figure}
    \centering
    \includegraphics[width=\linewidth]{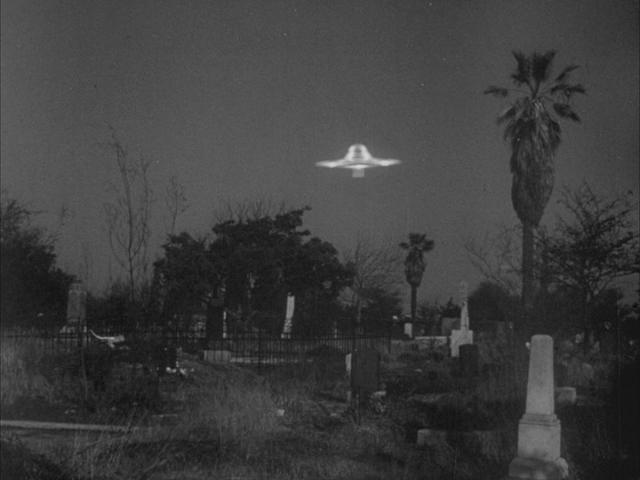}
    \caption{Still frame from Plan 9 from Outer Space (1959; dir. Ed Wood) detailing an invasion of Earth by extraterrestrial vampires and featuring a flying saucer that is totally not a tin plate attached to a string. Via Wikimedia Commons, marked public domain in the United States \citep{PlanNine02_WikimediaCommons}}
    \label{fig:plan9}
\end{figure}

For readers seeking a realistic depiction of an invasion by interstellar vampires, we recommend viewing Plan 9 from Outer Space by director Ed Wood (\citealt{Wood1959Plan9}, Fig.~\ref{fig:plan9}). 

Should future observations reveal a world that is sulfurous, suspiciously bright, and salty, the correct response is neither panic nor immediate submission to \textit{Nature}. First, run the retrieval again. If there is still something strange in the neighborhood, you know who to call.

\section{Acknowledgements}
No interstellar monsters were harmed in the creation of this work (although we may wish them harm). We acknowledge insightful discussions with Sebastian Zieba at the Center for Astrophysics | Harvard \& Smithsonian. His knowledge of the field helped improve the manuscript. The authors thank Mike Lund for the template and inspirational \textit{Acta Prima Aprilia} works. ChatGPT was used in this work with substantial human input. No ghosts were found in the machine.

%This research made use of Astropy,\footnote{http://www.astropy.org} a community-developed core Python package for Astronomy \citep{astropy:2013, astropy:2018}.

%This research has made use of NASA’s Astrophysics Data System.

%----------------------------------------------------------------------------------------
%	REFERENCE LIST
%----------------------------------------------------------------------------------------

\bibliographystyle{apalike}
\bibliography{main}

%----------------------------------------------------------------------------------------

%\end{multicols}

\end{document}